\documentclass[aps,prl,showpacs,twocolumn]{revtex4}

\usepackage{graphicx}

\def\cA{\mathcal{A}}
\def\vR{{\bf R}}
\def\vk{{\bf k}}
\def\vr{{\bf r}}
\def\vp{{\bf p}}

\def\pr{Phys. Rev.\ }
\def\jap{J. Appl. Phys.\ }

\begin{document}

%\begin{CJK*}{Bg5}{bsmi}

\title{Magnetic circular dichroism from the impurity band in III-V diluted magnetic semiconductors}
\author{Jian-Ming Tang
%(´ö°·»Ê)
}
\affiliation{Department of Physics, University of New Hampshire, Durham, NH 03824-3520}
\author{Michael E. Flatt\'e}
\affiliation{Department of Physics and Astronomy, University of Iowa, Iowa City, IA 52242-1479}

%\date{\today}

\begin{abstract}

The magnetic circular dichroism of III-V diluted magnetic semiconductors, calculated within a theoretical framework suitable for highly disordered materials, is shown to be dominated by optical transitions between the bulk bands and an impurity band formed from magnetic dopant states. The theoretical framework incorporates real-space Green's functions to properly incorporate spatial correlations in the disordered conduction band and valence band electronic structure, and includes extended and localized electronic states on an equal basis. Our findings reconcile unusual trends in the experimental magnetic circular dichroism in III-V DMSs with the antiferromagnetic $p$-$d$ exchange interaction between a magnetic dopant spin and its host. 

\end{abstract}


\maketitle
%\end{CJK*}

In III-V diluted magnetic semiconductors (DMSs) such as Ga$_{\rm 1-x}$Mn$_{\rm x}$As, formed by doping nonmagnetic host semiconductors such as GaAs with magnetic acceptors such as Mn, the magnetic properties are highly correlated with the electrical and optical properties \cite{ASL02,Samarth2004,MSS05,JSM+06}.  The effect of the added spin-polarized holes from the magnetic dopants on the relative optical absorption of right ($\sigma^+$) and left ($\sigma^-$) circularly polarized light (magnetic circular dichroism, or MCD) has been extensively explored to probe the interactions between the $d$ states of the local magnetic moments and the $p$ states of the host valence band \cite{AHT+98,BCM+99,SMT+99,SBT01,KWF+03,LWP+05,CZC+07}.
% which are expected to be antiferromagnetic as in II-VI DMSs such as Zn$_{\rm 1-x}$Mn$_{\rm x}$Se. 
Yet fundamental puzzles remain: the light polarization most absorbed in  Ga$_{\rm 1-x}$Mn$_{\rm x}$As, which determines the sign of the MCD signal, is different from that of Zn$_{\rm 1-x}$Mn$_{\rm x}$Se. This suggests that the $p$-$d$ exchange interaction in  Ga$_{\rm 1-x}$Mn$_{\rm x}$As is ferromagnetic \cite{LWP+05}, however in the dilute limit the interaction between a Mn spin and the GaAs valence band is known to be antiferromagnetic\cite{Schneider1987}. Refs.~\cite{SMT+99,DOM+01} argue the unexpected sign of the MCD signal can be reconciled with an antiferromagnetic $p$-$d$ interaction by considering a large shift of the Fermi level in the valence band due to doping (Moss-Burstein shift). However, the required Moss-Burstein shift is large ($\sim 100$~meV) and the MCD spectrum should have a pronounced doping dependence that includes changing sign at low doping \cite{SMT+99}, whereas the observed doping dependence of the dominant features of the MCD spectrum is weak \cite{BCM+99} and the unexpected sign of the MCD signal is present in low-doped, paramagnetic  Ga$_{\rm 1-x}$Mn$_{\rm x}$As (even for $x\sim 0.005$) \cite{LWP+05}.
Furthermore, none of the above treatments adequately treat the role of disorder in the optical transitions, whereas measurements indicate the carrier mean free path to be less than $1$~nm \cite{Sorensen2003}, comparable to the Fermi wavelength. Experimental evidence that the Fermi level in Ga$_{1-x}$Mn$_x$As lies in the impurity band rather than in the valence band \cite{VEVB+97,OKR+01,YSK+05,Sapega2005,Burch2005,Burch2006,ASA+08} also requires reconsideration of the Moss-Burstein shift.

Here we find that a proper consideration of the strong spatially-localized perturbation of the electronic structure in the valence band due to the Mn dopant, including both the bound state of the acceptor and the perturbations of the continuum states near the dopant, is essential to accurately calculate the magnetic circular dichroism. A key element of the successful calculation of these properties is our  approach to impurity averaging the optical absorption in magnetic semiconductors. Traditional approaches impurity average separately over the conduction band electronic structure and the valence band electronic structure, and then calculate the optical absorption for transitions between the two new effective bands as one would calculate optical absorption in a clean semiconductor. Instead we calculate the difference in optical absorption between the clean host and the host by calculating optical matrix elements between real-space Green's functions for a single dopant {\it before} impurity averaging. The effect of the short mean free path is included naturally by restricting the real-space sum to a small cluster of approximately that diameter. Thus optical transitions that do not conserve crystal momentum are included without artificially relaxing momentum conservation between impurity-averaged bands \cite{SBT01}.  

In addition to obtaining the correct sign of the MCD signal for low doping as well as high doping, our calculations identify the dominant transitions contributing to the MCD to be transitions between the bulk bands and the acceptor states bound to the Mn (that will form the impurity band). These transitions would be forbidden if impurity averaging and momentum conservation in optical absorption calculations were improperly imposed, as both initial and final states would have different crystal momentum. We further find that the amplitude of the MCD tracks the magnetization of the material, as seen experimentally in the temperature-dependence of the MCD \cite{BCM+99}.  Furthermore, the absorption onset is not sharp, and the MCD signal persists even for photon energies below the band gap of the host semiconductor, also as seen experimentally \cite{BCM+99,SMT+99}. 

We begin by describing the real-space tight-binding Green's function framework that permits the optical absorption of a single dopant to be calculated without impurity averaging.  The real-space Green's functions to be used have already been used in the calculation of the electronic structure of Mn dopants and have shown excellent agreement with experimental measurements of the local electronic structure near individual Mn dopants \cite{TF04,YSK+04b,YSK+07} and pairs of Mn \cite{YSK+05,KRT+06} embedded in GaAs.

\begin{figure}
\includegraphics[width=0.9\columnwidth]{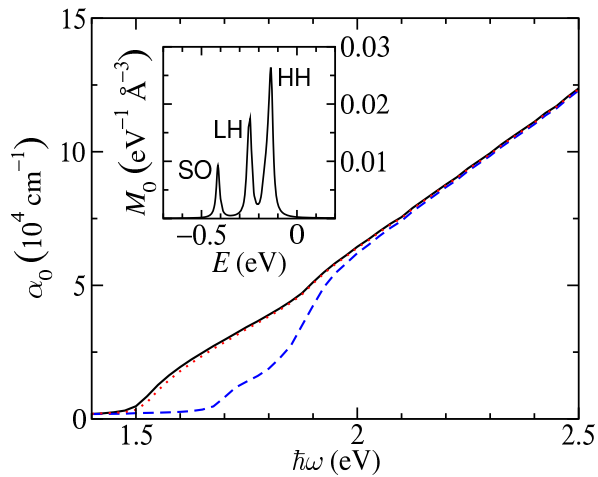}
\caption{(color online) Absorption coefficient ($\alpha_0$) for GaAs
  as function of photon energy ($\hbar\omega$) with the Fermi level at
  $0.1$ eV (solid line), $0$ eV (dotted line), and $-0.1$ eV (dashed
  line) relative to the valence-band maximum.  The step increase near
  $1.9$ eV shows the onset of the split-off band.  The inset shows an
  example of $M_0(E,E+\hbar\omega)$ as a function of $E$ with
  $\hbar\omega=2$ eV.  The three peaks from highest to lowest
  correspond to the cases where the initial states are in the
  heavy-hole (HH), the light-hole (LH), and the split-off (SO) valence bands. }
\label{fig:homo_abs}
\end{figure}

Following the formulation in Ref.~\onlinecite{YC01}  the absorption coefficients for circularly polarized light
($\sigma^\pm$) can be written
\begin{eqnarray}
  \alpha^\pm & = & \frac{\pi e^2}{n\varepsilon_0 mc\omega}\int dE f(E)\left[1-f(E^\prime)\right]M^\pm(E, E^\prime) \;,
\label{eq:absorption}
\end{eqnarray}
where $E$ is the initial state energy, $E^\prime=E+\hbar\omega$ the
final state energy, $\hbar\omega$ the photon energy, $e$ the
electron charge, $n$ the index of refraction, $\varepsilon_0$
the vacuum permittivity, $m$ the bare electron mass, $c$ the
speed of light in vacuum, $f(E)$ the Fermi-Dirac
function, and $M^\pm(E,E^\prime)$ the optical transition
strength per unit energy per unit volume.  In the electric dipole approximation
%$M^\pm(E,E^\prime)$ can be written as
\begin{eqnarray}
  M^\pm(E,E^\prime) & = & \frac{1}{mV}{\rm tr}\left[\hat{p}^\dagger_\pm\hat{\cA}(E^\prime)\hat{p}_\pm\hat{\cA}(E) \right] \;,
\label{eq:strength}
\end{eqnarray}
where $V$ is the system volume, $\hat{p}_\pm=\hat{p}_x\pm i\hat{p}_y$
are the momentum operators (the light is propagating along the $z$
direction), and the spectral function, $\hat{\cA}(E)$, is a
combination of retarded and advanced Green's functions,
\begin{eqnarray}
  \hat{\cA}(E) & = & \frac{i}{2\pi}\left[\hat{G}^R(E)-\hat{G}^A(E)\right] \;.
\end{eqnarray}
The trace in
Eq.(\ref{eq:strength}) is taken over a set of L\"owdin
orbitals, $\phi_{a,l}(\vr-\vR_{j,a})$, where $j$ labels the primitive
unit cells, $a$ labels the atomic sites within a unit cell,
$\vR_{j,a}$ is the position vector of an atomic site, and $l$ labels the
atomic orbitals at each site.  To simplify our calculations, we assume
the momentum matrix elements are nonzero only between two L\"owdin
orbitals located at the same site and are independent of the type of
the atom at that site. In the 16-band $sp^3$ tight-binding model we
are left with only one type of momentum matrix element,
\begin{eqnarray}
  \left\langle\phi_{a,p_x\uparrow}(\vr-\vR_{j,a})|\hat{p}_x|\phi_{a,s\uparrow}(\vr-\vR_{j,a})\right\rangle & = & iP \;.
\end{eqnarray}
With the above simplifications $P$ can be linked directly to the momentum matrix element
between the conduction and valence band Brillouin-zone-center states, known from {\it bulk} $\vk\cdot\vp$
theory. Therefore, no additional parameter in our
tight-binding framework needs to be introduced. The situation becomes more complex (and less empirically constrained) if  momentum matrix elements between
neighboring atomic sites are permitted to be non-zero. Using the tight-binding parameters in
Ref.~\onlinecite{Cha77}, the momentum matrix element between the
zone-center states is
$\langle\Gamma_{4v,x}|\hat{p}_x|\Gamma_{1c}\rangle=0.914\,iP$. Thus we
find $P^2/m=17.3$ eV by setting the zone-center momentum matrix
element to be the same as used in Ref.~\onlinecite{Rossler1984}. We calculate the tight-binding Green's
functions with an energy linewidth of $10$~meV as in
Ref.~\onlinecite{TF04}. Lastly, we use a constant index of refraction,
$n = 3.878$ for bulk GaAs at $2$~eV~\cite{Aspnes1983} for the energy
range ($1.2$-$2.5$~eV) shown in this Letter.

We first apply this method to calculate the absorption coefficient for
bulk GaAs.  In this case Eq.~(\ref{eq:strength}) can be evaluated
exactly in momentum space and the results, $\alpha^\pm=\alpha_0$ and
$M^\pm=M_0$, are shown in Fig.~\ref{fig:homo_abs}. A good agreement
with the experimental data \cite{Sturge1962} as a function of the
absorbed photon energy is found, although the calculated overall magnitude
is larger by $\sim 50$~\%, probably due to the slightly too large conduction band mass typical for tight-binding models of III-V semiconductors. The energy dependence near the absorption
edge has a square-root dependence with an Lorentzian tail because the
quasiparticles have a finite lifetime and excitons are not included in
our calculations. At higher energy the absorption's energy dependence is approximately
linear. The shoulder near $1.9$~eV is the onset for the split-off band
($0.36$~eV below the valence-band maximum). The absorption curves
(solid and dotted lines) are approximately the same as long as the
Fermi level is in the gap. If the Fermi level lies in the valence
band, the absorption edge moves to higher energy and the shape is
significantly altered. This shape change is not observed for
Ga$_{1-x}$Mn$_x$As, and thus adds support to interpretations placing the Fermi level in
the impurity band rather than in the valence band.

\begin{figure}
\includegraphics[width=0.9\columnwidth]{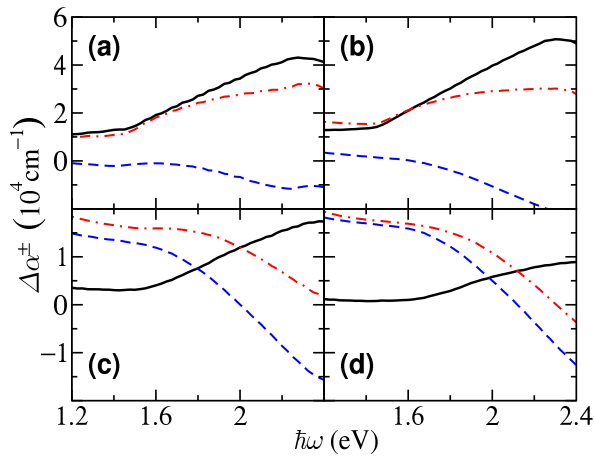}
\caption{(color online) Absorption coefficients, $\Delta\alpha^+$ (dashed line), $\Delta\alpha^-$ (dash-doted line), and $\alpha^- -\alpha^+$ (solid line), as functions of photon energy, $\hbar\omega$, with the Fermi level at (a) $0.1$~eV, (b) $0.05$~eV, (c) $0$~eV, and (d) $-0.1$~eV relative to the valence-band maximum.
}
\label{fig:MCD}
\end{figure}

In the presence of Mn dopants there are both localized states and
extended states. Eq.~(\ref{eq:strength}) is now evaluated in real
space by summing over the lattice sites within in a finite cluster. We
assume that the total MCD is the sum of MCD from these disjoint
clusters containing single Mn dopants. In doing so we have neglected
Mn-Mn interactions. This approximation is sufficient to describe structures in the optical absorption that are broader in energy than
the energy splittings due to Mn-Mn interactions ($\sim 0.1$ eV for a Mn-Mn
separation $\sim 1$ nm \cite{TF04}). Calculations based on disjoint clusters containing single Mn dopants automatically produce MCD features that are directly proportional to the material's magnetization, consistent with the behavior of the principal experimental MCD features~\cite{BCM+99}. The cluster size is sufficiently large to account for transitions involving the localized Mn
acceptor states or spin-polarized scattering resonances. The transitions between extended states exhibit finite-size effects  in these cluster sums, but those do not contribute to MCD, and can be more accurately captured by the momentum-space calculation of Fig.~\ref{fig:homo_abs}. Therefore, only the difference
$\Delta M^\pm = M^\pm - M_0$ is evaluated using finite clusters,
where $M_0$ refers to the results without Mn.

In our calculations, the light beam is parallel to one of the crystal
axes and is also parallel to the Mn magnetic moment. The model of Mn
is described in Ref.~\onlinecite{TF04}. The acceptor states are bound
mainly by the $p$-$d$ exchange interaction, which is described by a
spin-dependent potential ($V_{pd}$) present at the four
1st-nearest-neighbor sites. $V_{pd}=3.634$~eV \cite{YSK+07} is set to
obtain the experimental binding energy ($113$ meV). An on-site
potential ($V_n$) accounts for the direct Coulomb contribution to the
binding energy, and is chosen to be $1$~eV. Thus,
\begin{eqnarray}
V_{\rm Mn} & = & V_n\sum_{\ell,s} c^\dagger_{0,\ell,s}c_{0,\ell,s} + V_{pd}\!\!\!\!\!\sum_{\stackrel{n\in{\rm 1^{st}NN}}{\ell\in {p_x,p_y,p_z}}}\!\!\!\!\!c^\dagger_{n,\ell,\downarrow}c_{n,\ell,\downarrow} \;,
\end{eqnarray}
where $c^\dagger\left(c\right)$ is the creation (annihilation)
operator for electrons, $n$ labels atomic sites (Mn is at $n=0$),
$\ell$ labels atomic orbitals, and $s$ labels spins. The spin
quantization axis is parallel to the light propagation direction. The spin orientation of the Mn core $3d$ electrons is
antiparallel to the light propagation direction, whereas the spin of the acceptor states is parallel to the Mn core spin. Thus $V_{pd}>0$ represents an {\em
  antiferromagnetic} $p$-$d$ exchange interaction from the perspective
of holes.  The acceptor states are almost fully spin polarized and
split into three energy levels due to the spin-orbit interaction\cite{TF04}. As a
result, the three levels have quite distinct orbital-angular-momentum
character, and the top and bottom levels are coupled to opposite polarizations of
circularly-polarized light. The upper level (farthest from the valence
band edge) has orbital angular momentum parallel to spin. 

Our MCD results were obtained with a quasi-spherical cluster enclosing $99$ atoms, with one Mn atom in the center.
Up to the 8th nearest neighbors are included, which are $8$~\AA{} away from the
Mn.  $\sim 70\%$ of the acceptor state is included in this cluster, which  corresponds to approximately $2$\% Mn concentration.  The
results for different hole carrier concentrations are shown in
Fig.~\ref{fig:MCD} and we can see that the relative size of $\alpha^-$
is always larger than $\alpha^+$, no matter the doping. Note that we have ignored the
orbital mixing in the impurity band by Mn-Mn interaction, which would reduce the MCD, and the
energy broadening is taken into account only through a
$10$ meV linewidth in the single-particle Green's function. Therefore,
the overall calculated MCD magnitude we obtain is larger than experimentally measured.

\begin{figure}
\includegraphics[width=0.9\columnwidth]{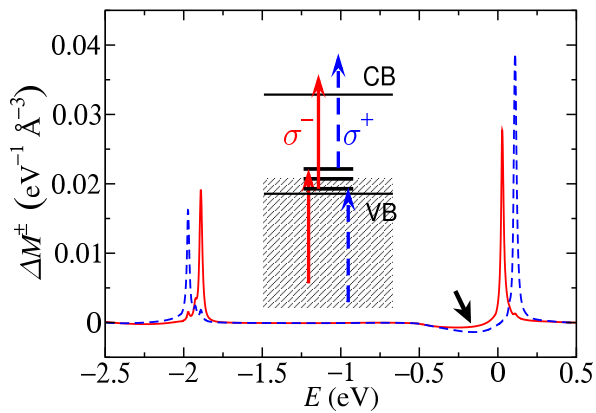}
\caption{ (color online) The optical transition strength, $\Delta M^+(E,E+\hbar\omega)$ (dashed line), and $\Delta M^-(E,E+\hbar\omega)$ (solid line), as functions of $E$ with $\hbar\omega=2$ eV. The four peaks correspond to the four distinct processes shown by the arrows in the energy diagram at the center. States in the shaded region are occupied if the Fermi level lies in the impurity band.  }
\label{fig:deltaM}
\end{figure}

\begin{figure}
\includegraphics[width=0.9\columnwidth]{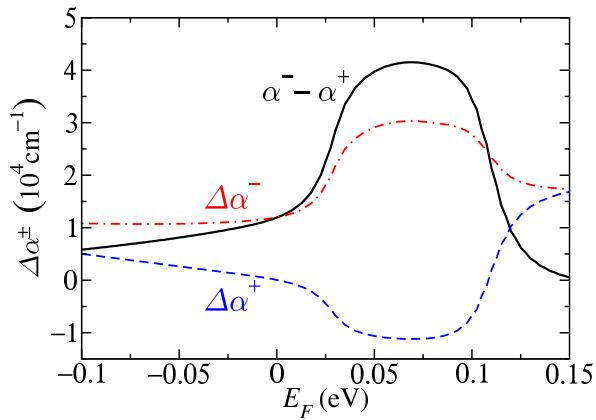}
\caption{ (color online) The absorption-coefficient differences,
  $\Delta\alpha^+$ (dashed line), $\Delta\alpha^-$ (dash-doted line),
  and $\alpha^- -\alpha^+$ (solid line), with $\hbar\omega=2$~eV as a
  function of the Fermi level relative to the valence-band maximum. }
\label{fig:MCD_dw2}
\end{figure}

To understand the nature of the doping-independent ``positive'' ($\alpha^->\alpha^+$)
MCD signal,  the contributions from different optical
transitions at the typical photon energy ($2$~eV) are shown in Fig.~\ref{fig:deltaM} as a function of the initial-state energy. Two peaks for each
circularly polarized light are present, corresponding to the processes
associated with the transition of an electron to or from the impurity levels. In our
configuration, the upper impurity level has predominately angular momentum
projection $-1$ and the bottom level has projection $+1$ because the
spin-orbit interaction favors the orbital angular momentum parallel to
spin (in the valence electron convention). The processes between the  valence and conduction
band are indicated by the short arrow in
Fig.~\ref{fig:deltaM}.  The sharpness of the peaks of $M$ in the
energy space comes from the small broadening factor ($10$~meV) that we
used for the impurity level. From Fig.~\ref{fig:deltaM} it is apparent that both
$\sigma^+$ processes are suppressed by the Pauli exclusion principle
if the Fermi level lies in the middle of the impurity band.  This is
further illustrated in Fig.~\ref{fig:MCD_dw2}.  We see that the
dichroism is the strongest when the Fermi level is between the
upper and the lower impurity levels.  The suppression of the positive
MCD signal when the Fermi level is above the impurity band is
consistent with recent observations in donor-compensated systems
\cite{CZC+07}.

In conclusion, a  theoretical approach to calculating the
optical properties of DMSs using real-space tight-binding Green's functions naturally explains several of the puzzling experimental findings that have been insufficiently explained using impurity-averaged conduction and valence states. The MCD measurements
agree with anti-ferromagnetic $p$-$d$ exchange
interaction expected from single-Mn measurements, and no Moss-Burstein shift is required to explain the sign of the MCD measurements. There is no sharp absorption onset due to transitions from the valence band to the impurity band and the dominant MCD features are proportional to the material magnetization, as previously reported from  experimental observations.

This work was supported in part by an ONR MURI and an ARO MURI.

%\bibliography{semiconductor}

\end{document}